\documentclass[graybox]{svmult}

\usepackage{mathptmx}
\usepackage{helvet}
\usepackage{courier}
\usepackage{type1cm}

\usepackage{makeidx}
\usepackage{graphicx}

\usepackage{multicol}
\usepackage[bottom]{footmisc}
\usepackage{url}

\usepackage{subcaption}
\usepackage{amsmath}

\makeindex

\begin{document}

\title*{Dengue in Madeira Island\thanks{This is a preprint of a paper
whose final and definite form will be published in the volume \emph{Mathematics of Planet Earth}
that initiates the book series \emph{CIM Series in Mathematical Sciences} (CIM-MS) published by Springer.
Submitted Oct/2013; Revised 16/July/2014 and 20/Sept/2014; Accepted 28/Sept/2014.}}

\author{Helena Sofia Rodrigues, M. Teresa T. Monteiro, Delfim F. M. Torres,
Ana Clara Silva, Carla Sousa and Cl\'{a}udia Concei\c{c}\~{a}o}

\authorrunning{H. S. Rodrigues et al.}

\institute{Helena Sofia Rodrigues
\at  Escola Superior de Ci\^{e}ncias Empresariais, Instituto Polit\'{e}cnico de Viana do Castelo, Portugal\\
\email{sofiarodrigues@esce.ipvc.pt}
\and M. Teresa T. Monteiro
\at ALGORITMI, Departamento de Produ\c{c}\~{a}o e Sistemas, Universidade do Minho, Portugal\\
\email{tm@dps.uminho.pt}
\and Delfim F. M. Torres
\at CIDMA, Departamento de Matem\'{a}tica, Universidade de Aveiro, Portugal\\
\email{delfim@ua.pt}
\and Ana Clara Silva
\at Instituto de Administra\c{c}\~{a}o da Sa\'{u}de e Assuntos Sociais, IP-RAM, Portugal\\
\email{anaclarasilv@gmail.com}
\and Carla Sousa
\at Instituto de Higiene e Medicina Tropical, Universidade Nova de Lisboa, Portugal\\
\email{CASousa@ihmt.unl.pt}
\and Cl\'{a}udia Concei\c{c}\~{a}o
\at Instituto de Higiene e Medicina Tropical, Universidade Nova de Lisboa, Portugal\\
\email{claudiaconceicao@ihmt.unl.pt}}

\maketitle


\abstract{Dengue is a vector-borne disease and 40\%
of world population is at risk. Dengue transcends
international borders and can be found in tropical
and subtropical regions around the world,
predominantly in urban and semi-urban areas.
A model for dengue disease transmission,
composed by mutually-exclusive compartments representing
the human and vector dynamics, is presented in this study.
The data is from Madeira, a Portuguese island, where an unprecedented
outbreak was detected on October 2012. The aim of this work is to simulate
the repercussions of the control measures in the fight of the disease.}


\section{Introduction}

During the last decades, the global prevalence of dengue increased considerably.
Madeira's dengue outbreak of 2012 is the first epidemics in Europe since the
one recorded in Greece in 1928 \cite{Halstead1980}. Local transmission was also
reported, for the first time, in France and Croatia in 2010
\cite{Gjenero-Margan2011,Ruche2010} and the threat of possible
outbreaks of dengue fever in Europe is increasing.
According to a recent study \cite{Bhatt2013},
390 million dengue infections occur per year worldwide,
of which 96 million with clinical symptoms. Methods considered
by authorities for disease prevention include educational
and vaccination campaigns, preventive drugs
administration and surveillance programs.

Mathematical modeling plays a fundamental role in the study
of the evolution of infectious diseases \cite{Anderson,MyID:305,MyID:271}.
When formulating a model for a particular disease, a trade-off
between simple and complex models is always present. The former, omit several details
and are generally used for short-term and specific situations, but have the
disadvantage of possibly being naive and unrealistic. The complex models
have more details and are more realistic, but are generally more difficult
to solve and analyze or may contain parameters whose estimates
cannot be obtained \cite{Diekmann}. Here we are interested
in a dengue model defined by a system of ordinary differential equations,
which enables the evaluation of the infectious disease transmission patterns.

The text is organized as follows.
Section~\ref{Sec:2} presents some details about dengue,
such as disease symptoms and vector transmission issues.
The outbreak on Madeira island and measures to
fight against the epidemics are described in Section~\ref{Sec:3}.
In Section~\ref{Sec:4} the mathematical model for the interaction
between humans and mosquitoes is formulated, while numerical
experiments using distinct levels of control
are presented in Section~\ref{Sec:5}.
We end with Section~\ref{Sec:6}
of conclusions and ideas for future work.


\section{Dengue and the \emph{Aedes} mosquito}
\label{Sec:2}

Dengue is a vector-borne disease transmitted from an infected human to
an \emph{Aedes} mosquito, commonly \emph{Aedes aegypti} or \emph{Aedes albopictus},
during a female blood-meal \cite{Cattand2006}. Then, the infectious mosquito,
that needs regular meals of blood to mature their eggs, bites a potential healthy human
and transmits the disease, thus completing the extrinsic cycle of the virus.
Four dengue serotypes are known, designated as DEN-1, 2, 3 and 4, which cause
a wide spectrum of human disease, from asymptomatic cases to classic dengue
fever (DF) and more severe cases, known as dengue hemorrhagic fever (DHF).
Symptoms include fever, headache, nausea, vomiting, rash, and pain in the eyes,
joints, and muscles. Symptoms may appear up to two weeks after the bite of an
infected mosquito and usually last for one week. In severe cases, symptoms
may include intense stomach pain, repeated vomiting, and bleeding from the nose or gums
and can lead to death. Recovery from infection by one virus
provides lifelong immunity against that virus but only confers partial and transient
protection against subsequent infection by the other three serotypes. There is good
evidence that a sequential infection increases the risk of developing DHF \cite{Wearing2006}.

Unfortunately, there is no specific treatment for dengue. Activities,
such as triage and management, are critical in determining the clinical outcome of dengue.
A rapid and efficient front-line response
not only reduces the number of unnecessary hospital admissions but also saves lives.
Although there is no effective and safe vaccine for dengue, a number of candidates are
undergoing various phases of clinical trials \cite{Who2009}. With four closely
related serotypes that can cause the disease, there is a need for an effective vaccine that
would immunize against all four types; if not, a secondary infection could,
theoretically, lead to a DHF case. Another difficulty in the
vaccine production is that there is a limited understanding of how the disease typically
behaves and how the virus interacts with the immune system. Research to develop
a vaccine is ongoing and the incentives to study the
mechanism of protective immunity are gaining more support, now that
the number of outbreaks around the world is increasing
\cite{DengueNet}. Several mathematical models, including a few taking into account
vaccination and optimal control, have been proposed in the literature:
see \cite{MyID:168,Sofia2013,MyID:263,MyID:283,MyID:306} and references therein.

The life cycle of the mosquito has four distinct stages: egg, larva, pupa and adult.
The first three stages take place in water, whilst air is the medium
for the adult stage. \emph{Aedes} females have a peculiar oviposition behavior:
they do not lay all the eggs of an oviposition at once, in the same breeding site,
but rather release them in different places, thus increasing the probability
of successful births \cite{Otero2008,Shuman}.
In urban areas, \emph{Aedes aegypti} breed on water collections in artificial containers
such as cans, plastic cups, used tires, broken bottles and
flower pots. With increasing urbanization and crowded cities,
environmental conditions foster the spread of the disease that,
even in the absence of fatal forms, breed significant economic and
social costs (absenteeism, immobilization,
debilitation and medication) \cite{Derouich2006}.

It is very difficult to control or eliminate
\emph{Aedes} mosquitoes because they are highly resilient,
quickly adapting to changes in the environment and they have the
ability to rapidly bounce back to initial numbers after disturbances
resulting from natural phenomena (e.g., droughts) or human interventions
(e.g., control measures). We can safely expect that transmission thresholds
will vary depending on a range of factors.
Reduction of vector populations, both adult mosquitoes
and in immature states, is currently the only way to prevent dengue.


\section{Madeira's dengue outbreak}
\label{Sec:3}

\begin{figure}
\center
\sidecaption
\includegraphics[scale=0.5]{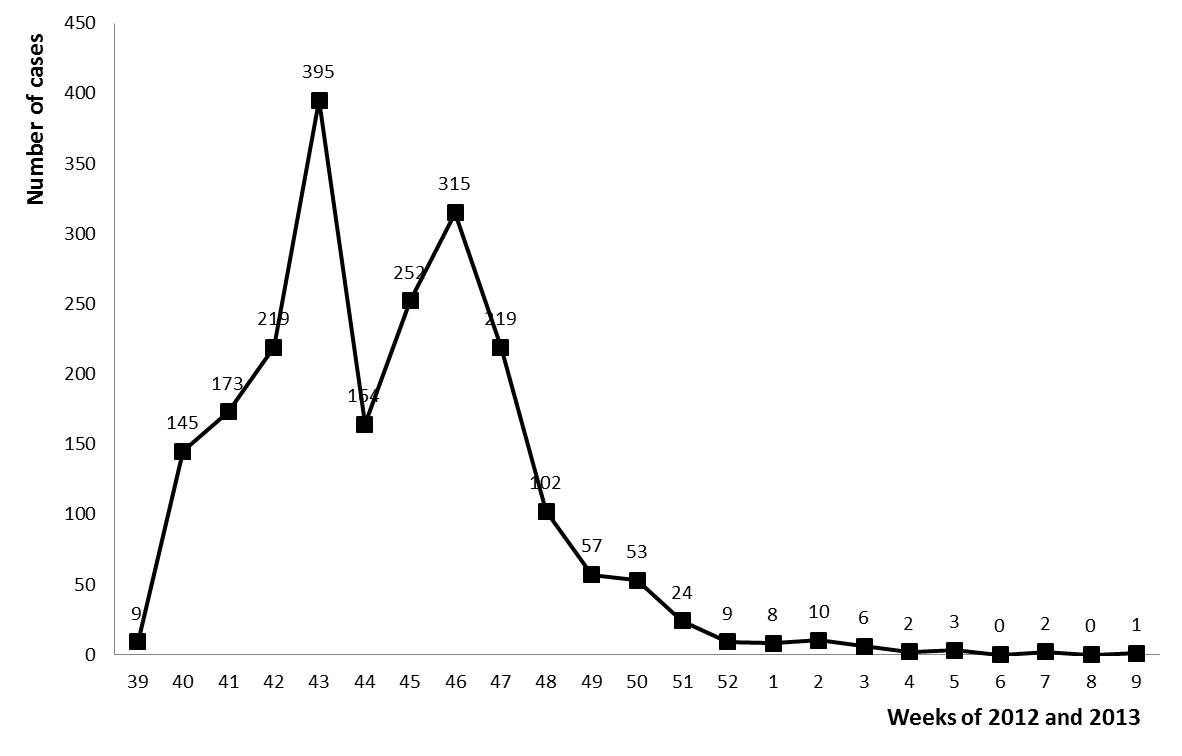}
\caption{Notified dengue fever cases in Madeira, by week,
from October 2012 to February 2013 (Source: Instituto de Administra\c{c}\~{a}o da Sa\'{u}de
e Assuntos Sociais, Regi\~{a}o Aut\'{o}noma da Madeira)}
\label{dengue_by_week}
\end{figure}

An outbreak of dengue fever, that lasted about 21 weeks between early October 2012 and
late February 2013, occurred in Madeira, a Portuguese island, whose capital is Funchal.
As March 12th, 2013, 2168 probable cases of dengue fever have been reported, of which 1084 were
laboratory confirmed. All reported cases refer to the resident population of the island and no
deaths or severe cases were reported. On the same day, according to the data available,
the outbreak was considered finished by the Portuguese Health
Authorities, since there was no autochthonous cases in the island \cite{DGS}.
The notified dengue fever cases in Madeira, by week, are in Figure~\ref{dengue_by_week}.
Note that the number of confirmed dengue cases is lower than the notified ones.

\begin{figure}
\center
\sidecaption
\includegraphics[scale=0.5]{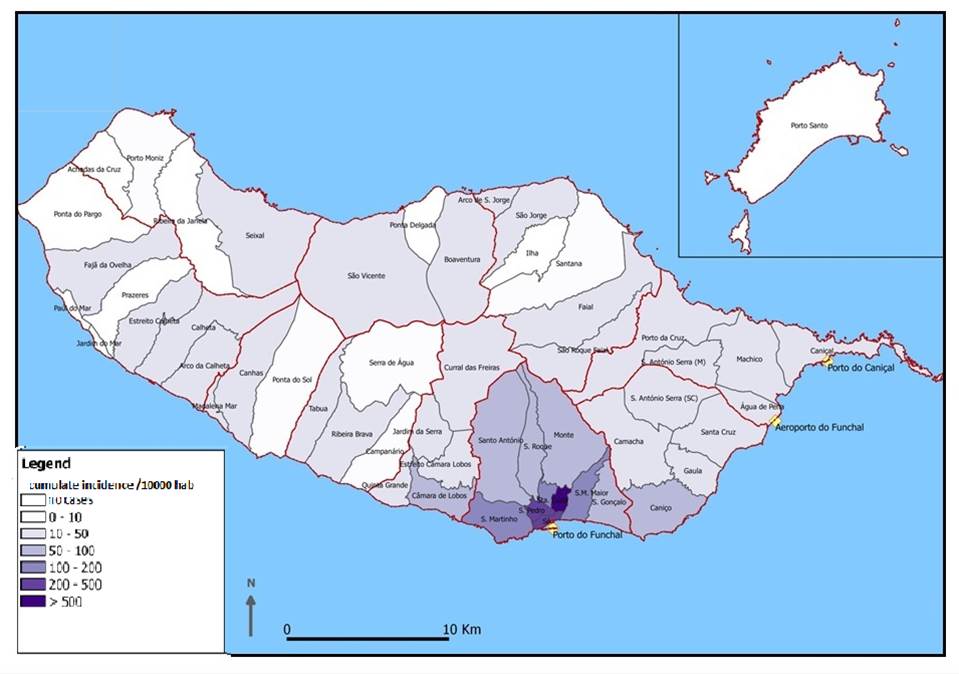}
\caption{Cumulative incidence of dengue in Madeira, by parish,
from October 2012 to February 2013 (Source: Instituto de Administra\c{c}\~{a}o da Sa\'{u}de
e Assuntos Sociais, Regi\~{a}o Aut\'{o}noma da Madeira)}
\label{Madeira_parish}
\end{figure}

In Figure~\ref{Madeira_parish} it is possible to see
the cumulative incidence of dengue cases
along the island, by parish. Santa Luzia parish
is the one that recorded the highest proportion of patients.
As the mosquito lives mainly in urban areas with high human population density,
and the vast majority of human cases were observed in this civil parish of the Funchal council,
our study is constrained to this area.

The mosquito \emph{Aedes aegypt} was detected in Madeira, for the first time, in 2005.
The National Institute of Health Doutor Ricardo Jorge (INSA) performs
reference laboratory diagnosis of dengue in Portugal. INSA conducted
confirmatory laboratory diagnosis and identified the presence of
DEN-1 virus in human samples \cite{INSA}.
Molecular analyses reported that the virus in Madeira island could have origin
in Brazil or Venezuela, where the virus presents similar features and with whom
there are intensive movements of trade and people \cite{INSA}.

After the acknowledgement of the presence of the dengue mosquito in Madeira,
local Health authorities implemented several strategies to
control this invasive specie of mosquitoes. However, the results showed small effects.
\emph{Aedes aegypti} in Madeira present a high resistance level to DDT, permethrin
and deltamethrinm, the common tools allowed by the
World Health Organization (WHO) \cite{tese_goncalo}. Therefore,
local measures changed to educational campaigns and entomological surveillance,
to monitor the vector spread using traps, both for eggs and adult forms.
Educational campaigns appealed the population to apply
repellent and wear large clothes to avoid mosquito bites.
Moreover, all recipients that could serve to breed the mosquito,
like water collections in artificial containers (e.g., cans, plastic cups,
used tires, broken bottles and flower pots), were asked to be removed or covered.
Media-based tools were used to inform the population. These included newspapers,
TV programs, TV spots, radio programs, radio spots, flyers, internet sites,
announcements and specific talks in public places. A medical appointment dedicated
to the dengue disease was also implemented in a health unit in Funchal,
and a program for the monitoring of traps implemented. The number of eggs per
trap, dispersedly placed along the island, with emphasis on the southern
slope, were counted in order to understand their spatial distribution.
Weekly entomological reports of \emph{Aedes aegypti} in Madeira island were,
and still are, broadcasted to sectorial partners. The application of insecticide
was only applied in strategic places, such as the central hospital, the health unit
dedicated to the attendance of dengue cases and a school identified as a
transmission area \cite{carla2012}.


\section{The mathematical model}
\label{Sec:4}

Taking into account the model presented in
\cite{Dumont2010,Dumont2008} and the considerations
of \cite{Sofia2009,Sofia2010c}, a temporal mathematical model
to study Madeira's dengue outbreak is here proposed.
It includes three epidemiological states for humans:
\begin{quote}
\begin{tabular}{ll}
$S_h(t)$ & --- susceptible (individuals who can contract the disease);\\
$I_h(t)$ & --- infected (individuals who can transmit the disease);\\
$R_h(t)$ & --- resistant (individuals who have been infected and have recovered).
\end{tabular}
\end{quote}
These compartments are mutually-exclusive. There are three other state variables, related
to the female mosquitoes (male mosquitos are not considered because they do not bite
humans and consequently do not influence the dynamics of the disease):
\begin{quote}
\begin{tabular}{ll}
$A_m(t)$ & --- aquatic phase (includes egg, larva and pupa stages);\\
$S_m(t)$ & --- susceptible (mosquitoes that can contract the disease);\\
$I_m(t)$ & --- infected (mosquitoes that can transmit the disease).
\end{tabular}
\end{quote}
In order to make a trade-off between simplicity and reality of the epidemiological model,
some assumptions are considered:
\begin{itemize}
\item there is no vertical transmission, \emph{i.e.},
an infected mosquito cannot transmit the disease to their eggs;

\item total human population $N_h$
is constant: $S_h(t)+I_h(t)+R_h(t) = N_h$ at any time $t$;

\item the population is homogeneous, which means that every individual
of a compartment is homogeneously mixed with the other individuals;

\item immigration and emigration are not considered during the period under study;

\item homogeneity between host and vector populations, that is,
each vector has an equal probability to bite any host;

\item humans and mosquitoes are assumed to be born susceptible.
\end{itemize}
To analyze the disease evolution, two control measures are considered in the model:
\begin{quote}
\label{eq:star:3}
\begin{tabular}{ll}
$c_{m}(t)$ & --- proportion of insecticide (adulticide), $0 \leq c_m(t) \leq1$;\\
$1-\alpha(t)$ & --- proportion of ecological control, $0 < \alpha(t) \leq 1$.
\end{tabular}
\end{quote}
The application of adulticides is the most common control measure.
However, its efficacy is often constrained by the difficulty in achieving sufficiently
high coverage of resting surfaces and the insecticide resistance by the mosquito.
Besides, the long term use of adulticide comports several risks: it can affect other species,
it is linked to numerous adverse health effects, including the worsening of asthma and respiratory problems.
The purpose of ecological control, that is, educational campaigns,
is to reduce the number of larval habitat areas available to mosquitoes.
The mosquitoes are most easily controlled by treating,
cleaning and/or emptying containers that
hold water, since the eggs of the specie are laid in water-holding
containers. The ecological control must be done by both public health
officials and residents in the affected areas. The participation
of the entire population in removing still water from domestic recipients
and eliminating possible breeding sites is essential \cite{Who2009}.

\begin{figure}[ptbh]
\begin{center}
\includegraphics[scale=0.5]{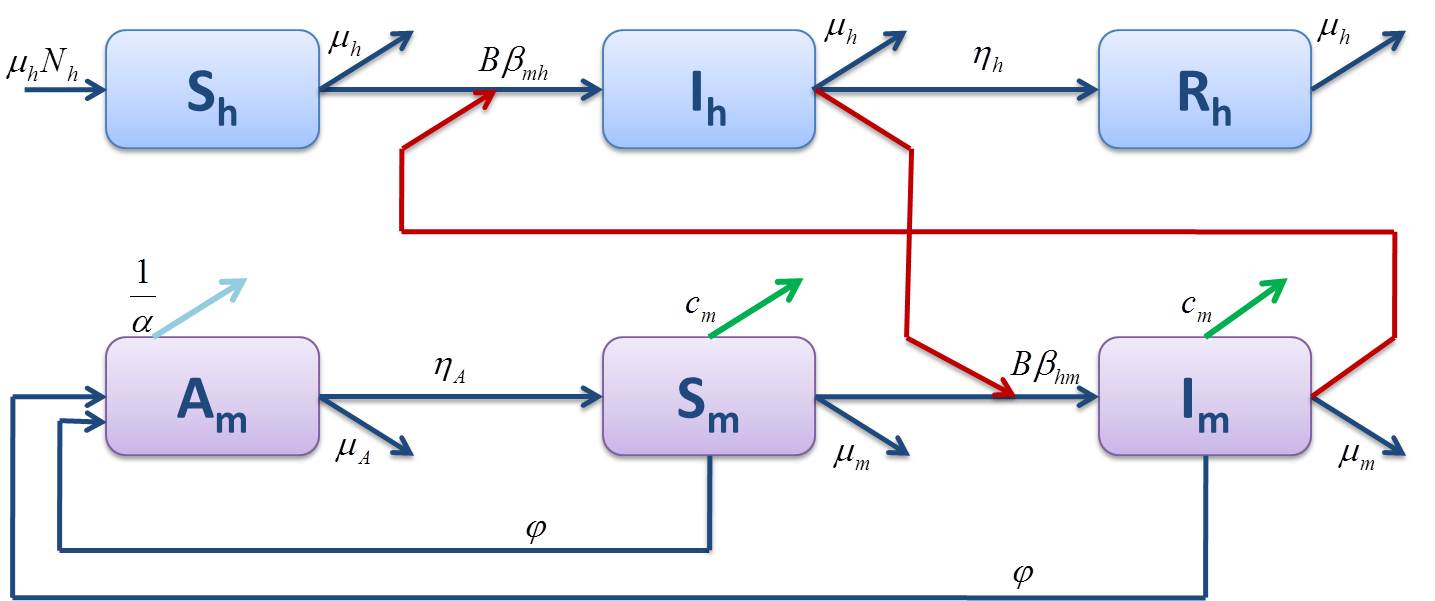}
\end{center}
\caption{Epidemiological model SIR \eqref{cap6_ode1}
+ ASI \eqref{cap6_ode2} \label{cap6_model}}
\end{figure}

Our dengue epidemic model makes use of the parameters described in Table~\ref{table_1}
\begin{table}[ptbh]
\begin{center}
\caption{Parameters in the epidemiological model \eqref{cap6_ode1}--\eqref{cap6_ode2}}
\label{table_1}
\begin{tabular}{lllll}
\hline
Para\- & Description & Range of values  & Value & Source\\
meter& & in literature & used & \\
\hline
$N_h $ & total population & & 112000 & \cite{censos2011}\\
$B$ & average daily biting (per day)& & 1/3 & \cite{Focks2000}\\
$\beta_{mh}$ & transmission probability from $I_m$ (per bite)& [0.25, 0.33] & 0.25 & \cite{Focks2000} \\
$\beta_{hm}$ & transmission probability from $I_h$ (per bite)& [0.25, 0.33] & 0.25 & \cite{Focks2000} \\
$1/\mu_{h}$ & average lifespan of humans (in days)& & $\displaystyle\frac{1}{79\times365}$ & \cite{censos2011} \\
$1/\eta_{h}$ & average viremic period (in days)& [1/15, 1/4] & 1/7 & \cite{Chan2012}\\
$1/\mu_{m}$ & average lifespan of adult mosquitoes (in days)& [1/45, 1/8]& 1/15 & \cite{Focks1993,Harrington2001,Freitas2007} \\
$\varphi$ & number of eggs at each deposit per capita (per day)& & 6 & \cite{Sofia2012}\\
$1/\mu_{A}$ & natural mortality of larvae (per day)& & 0.2363 & \cite{Barrios2013}\\
$\eta_{A}$ & maturation rate from larvae to adult (per day)& [1/11, 1/7]& 1/9 & \cite{Padmanabla2011}\\
$k$ & number of larvae per human & & 0.9 & \cite{Focks1995,Watson1999}\\
\hline
\end{tabular}
\end{center}
\end{table}
and consists of the system of differential equations
\begin{equation}
\label{cap6_ode1}
\begin{cases}
\displaystyle\frac{dS_h(t)}{dt}
= \mu_h N_h - \left(B\beta_{mh}\frac{I_m(t)}{N_h}+\mu_h\right)S_h(t)\\
\displaystyle\frac{dI_h(t)}{dt} = B\beta_{mh}\frac{I_m(t)}{N_h}S_h(t) -(\eta_h+\mu_h) I_h(t)\\
\displaystyle\frac{dR_h(t)}{dt} = \eta_h I_h(t) - \mu_h R_h(t)
\end{cases}
\end{equation}
coupled with the nonlinear control system
\begin{equation}
\label{cap6_ode2}
\begin{cases}
\displaystyle\frac{dA_m(t)}{dt}
= \varphi \left(1-\frac{A_m(t)}{\alpha(t) k N_h}\right)(S_m(t)+I_m(t))
-\left(\eta_A+\mu_A \right) A_m(t)\\
\displaystyle\frac{dS_m(t)}{dt} = \eta_A A_m(t)
-\left(B \beta_{hm}\frac{I_h(t)}{N_h}+\mu_m + c_m(t)\right) S_m(t)\\
\displaystyle\frac{dI_m(t)}{dt} = B \beta_{hm}\frac{I_h(t)}{N_h}S_m(t)
-\left(\mu_m + c_m(t)\right) I_m(t)
\end{cases}
\end{equation}
subject to the initial conditions
\begin{gather*}
S_h(0)=S_{h0}, \quad  I_h(0)=I_{h0}, \quad R_h(0)=R_{h0},\\
A_m(0)=A_{m0}, \quad S_{m}(0)=S_{m0}, \quad I_m(0)=I_{m0}.
\end{gather*}
Note that the differential equation related to the aquatic phase does not involve the control variable $c_m$,
because the adulticide does not produce effects in this stage of mosquito life.
Figure~\ref{cap6_model} shows a scheme of the model.

In the next section we study the reality of Madeira's outbreak,
using the most reliable information about the mosquito and the infected people.
For a mathematical analysis of the model, in particular the analysis of equilibrium points
and the basic reproduction number, we refer the reader to \cite{Sofia2013}.


\section{Numerical experiments}
\label{Sec:5}

\begin{figure}
\centering
\begin{subfigure}[b]{0.48\textwidth}
\centering
\includegraphics[width=\textwidth]{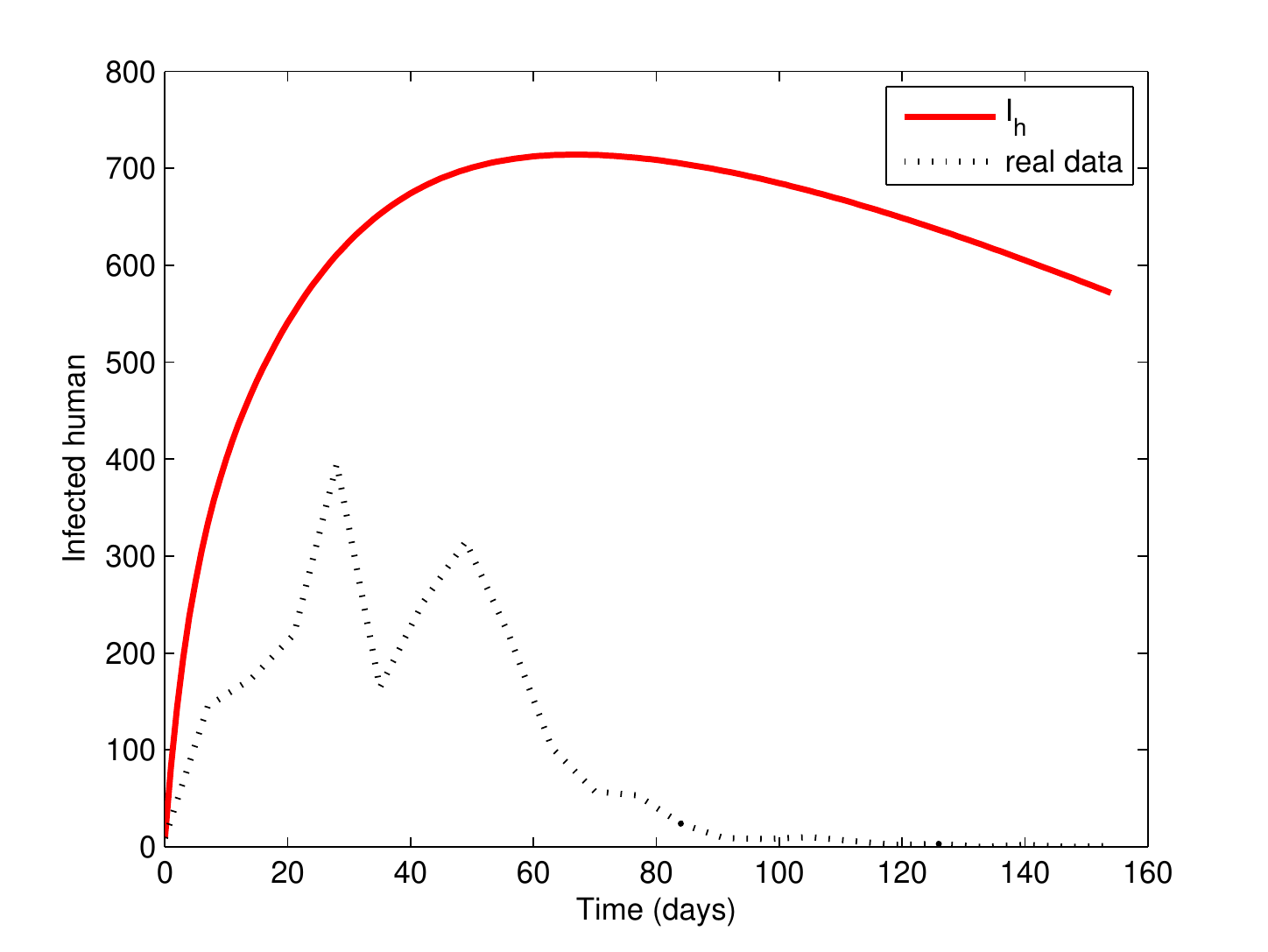}
\caption{(A) No educational campaigns, i.e. $1-\alpha \equiv 0$}
\label{madeira_sem_controlo}
\end{subfigure}
\begin{subfigure}[b]{0.48\textwidth}
\centering
\includegraphics[width=\textwidth]{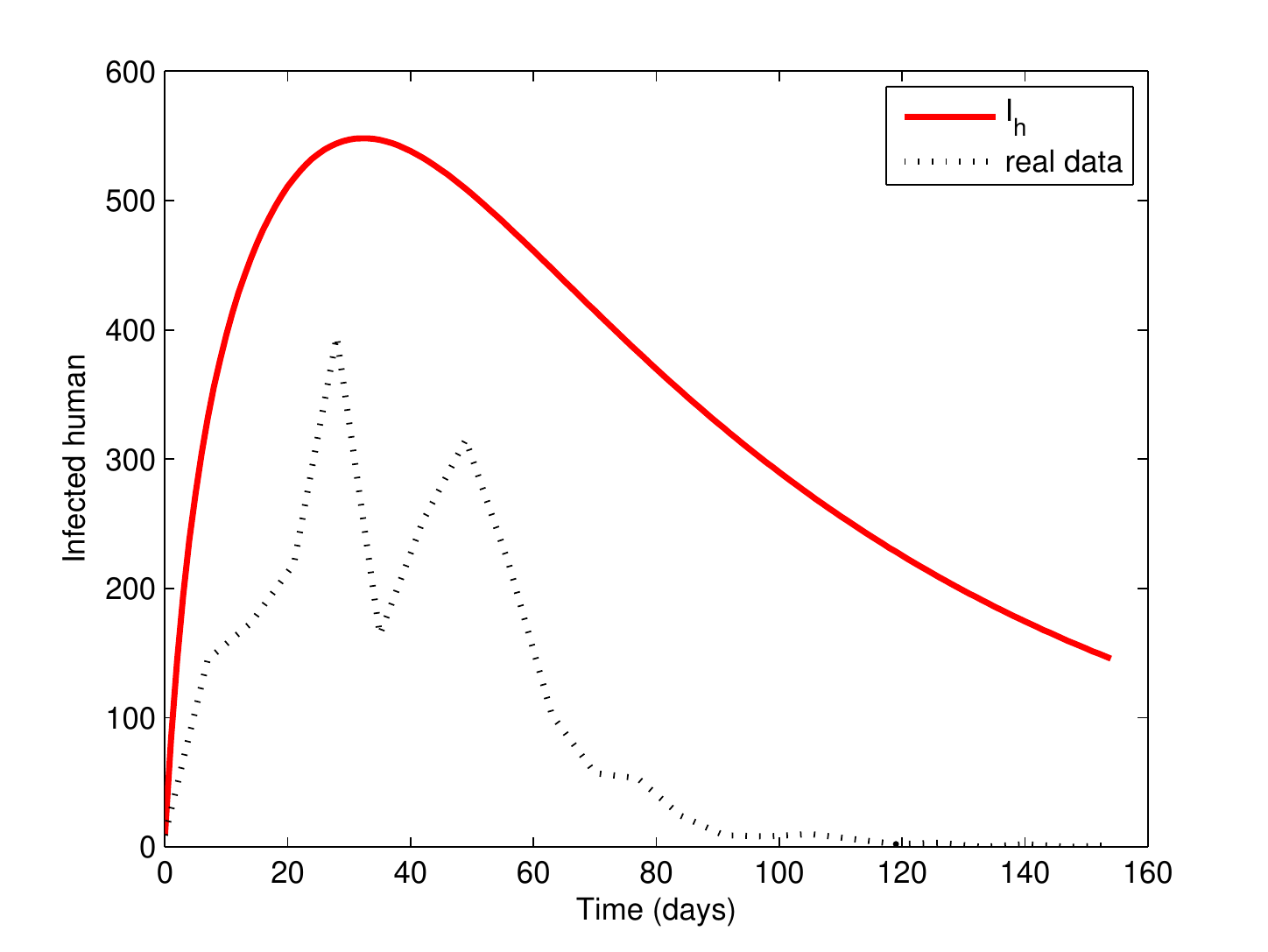}
\caption{(B) $1-\alpha \equiv 0.25$}
\label{madeira_alpha_25}
\end{subfigure}
\\
\begin{subfigure}[b]{0.48\textwidth}
\centering
\includegraphics[width=\textwidth]{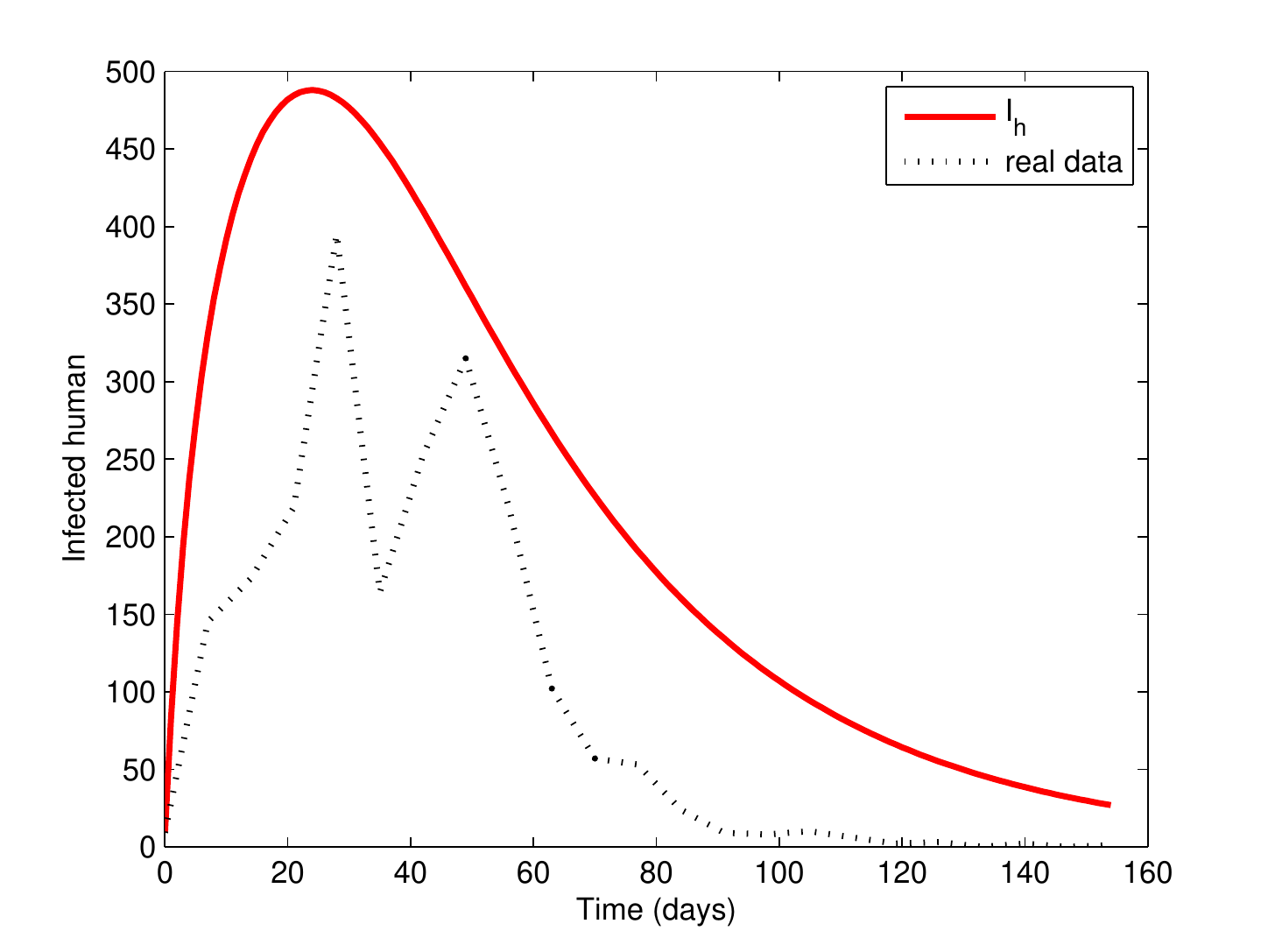}
\caption{(C) $1-\alpha \equiv 0.5$}
\label{madeira_alpha_50}
\end{subfigure}
\begin{subfigure}[b]{0.48\textwidth}
\centering
\includegraphics[width=\textwidth]{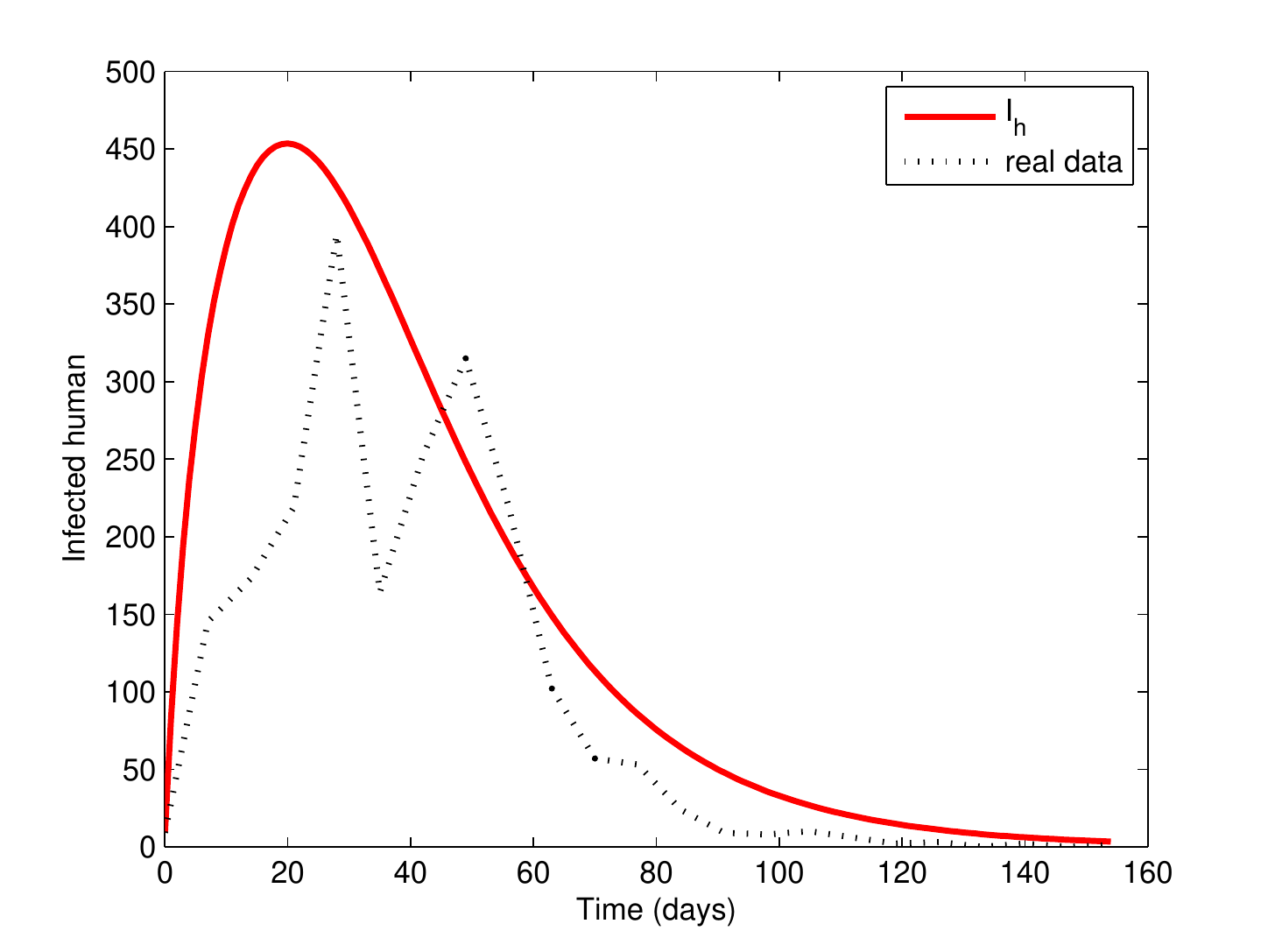}
\caption{(D) $1-\alpha \equiv 0.75$}
\label{madeira_alpha_75}
\end{subfigure}
\caption{Number of infected individuals
without insecticide usage (i.e., $c_m=0$) but with distinct
levels of educational campaigns (continuous line) \emph{versus}
observed real data (dotted line)}\label{educational}
\end{figure}

\begin{figure}
\centering
\begin{subfigure}[b]{0.48\textwidth}
\centering
\includegraphics[width=\textwidth]{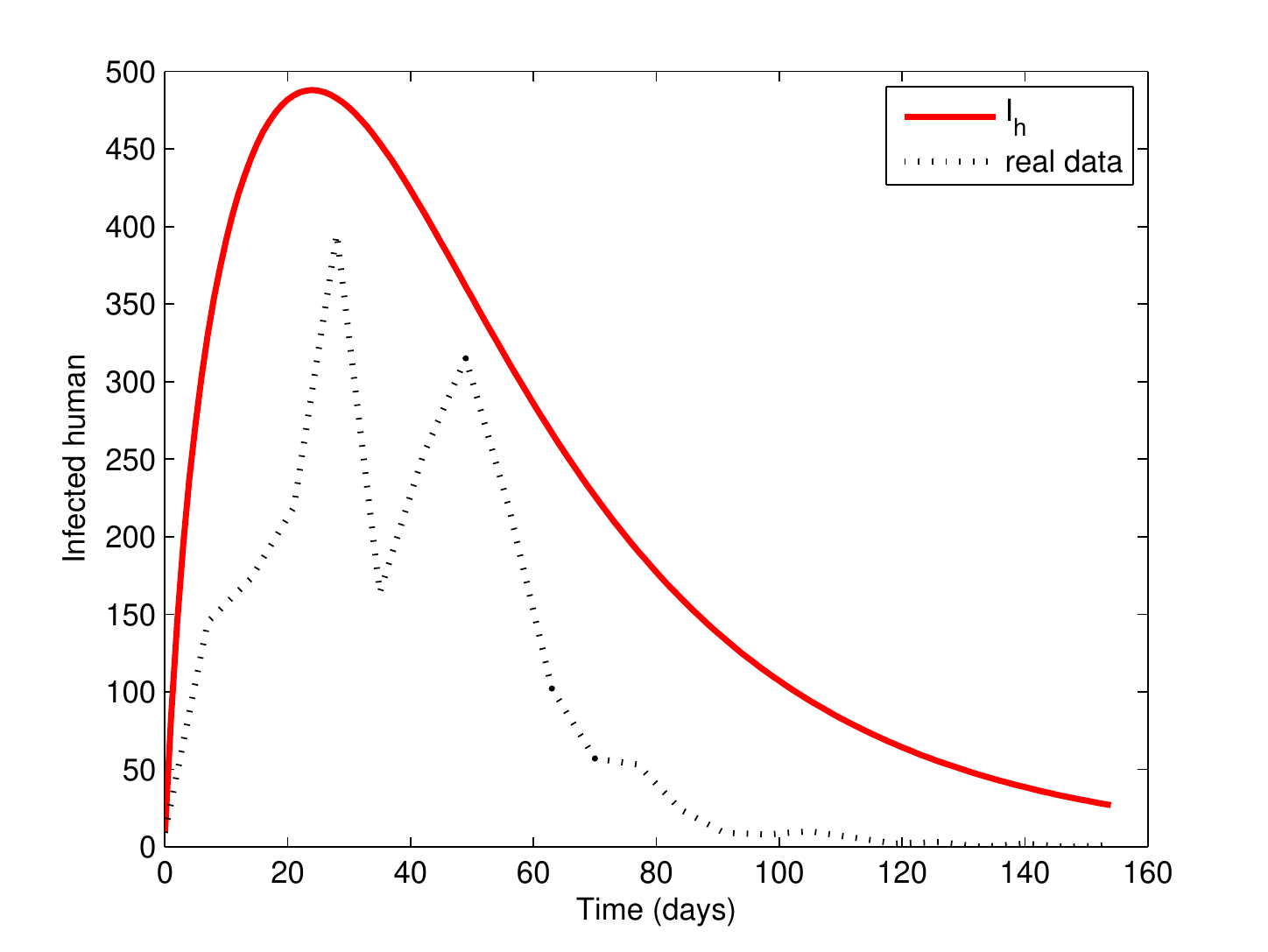}
\caption{(C) $c_m \equiv 0$}
\label{madeira_insecticida_0}
\end{subfigure}
\begin{subfigure}[b]{0.48\textwidth}
\centering
\includegraphics[width=\textwidth]{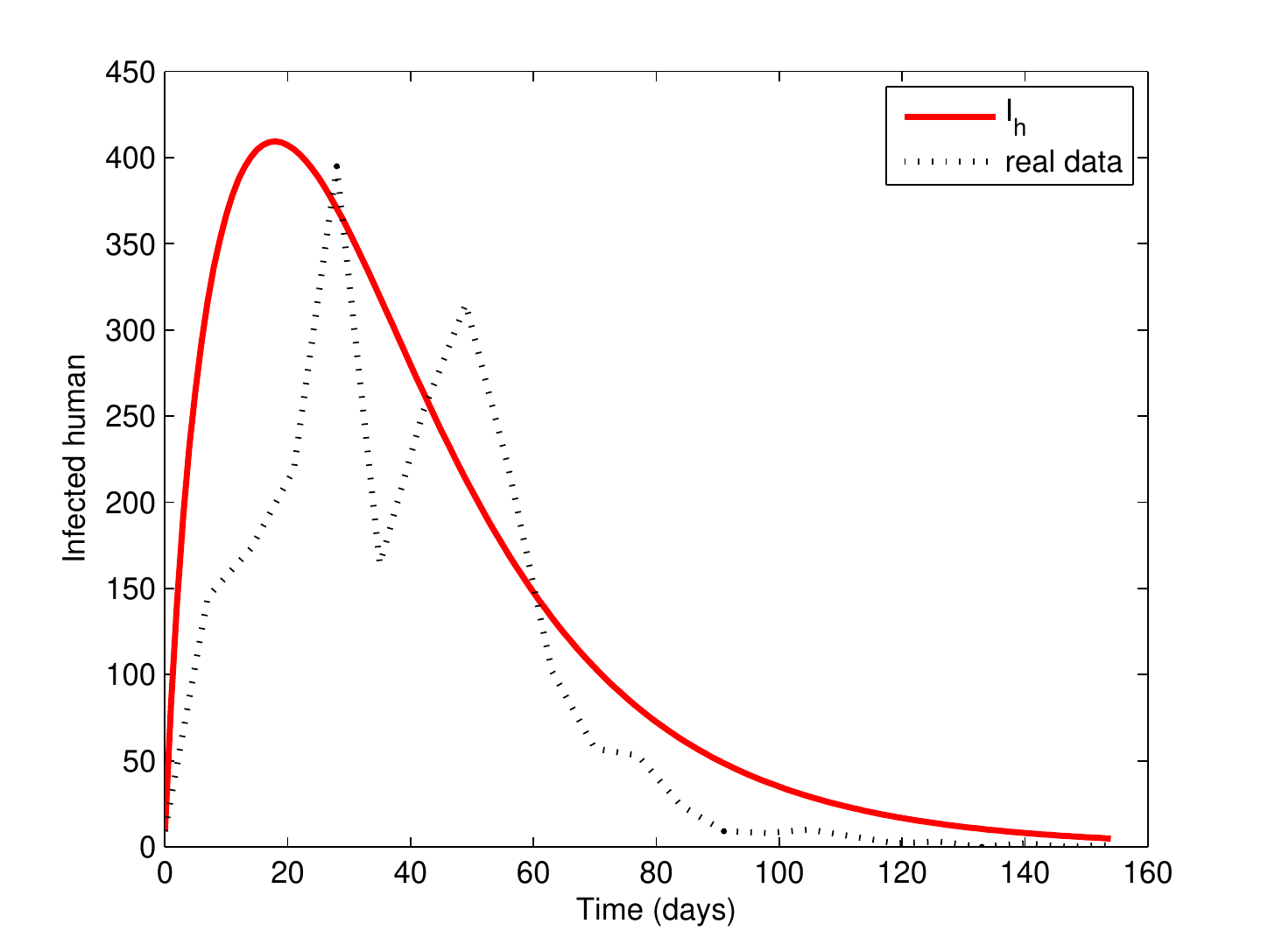}
\caption{(E) $c_m \equiv 0.01$}
\label{madeira_insecticida_1}
\end{subfigure}
\\
\begin{subfigure}[b]{0.48\textwidth}
\centering
\includegraphics[width=\textwidth]{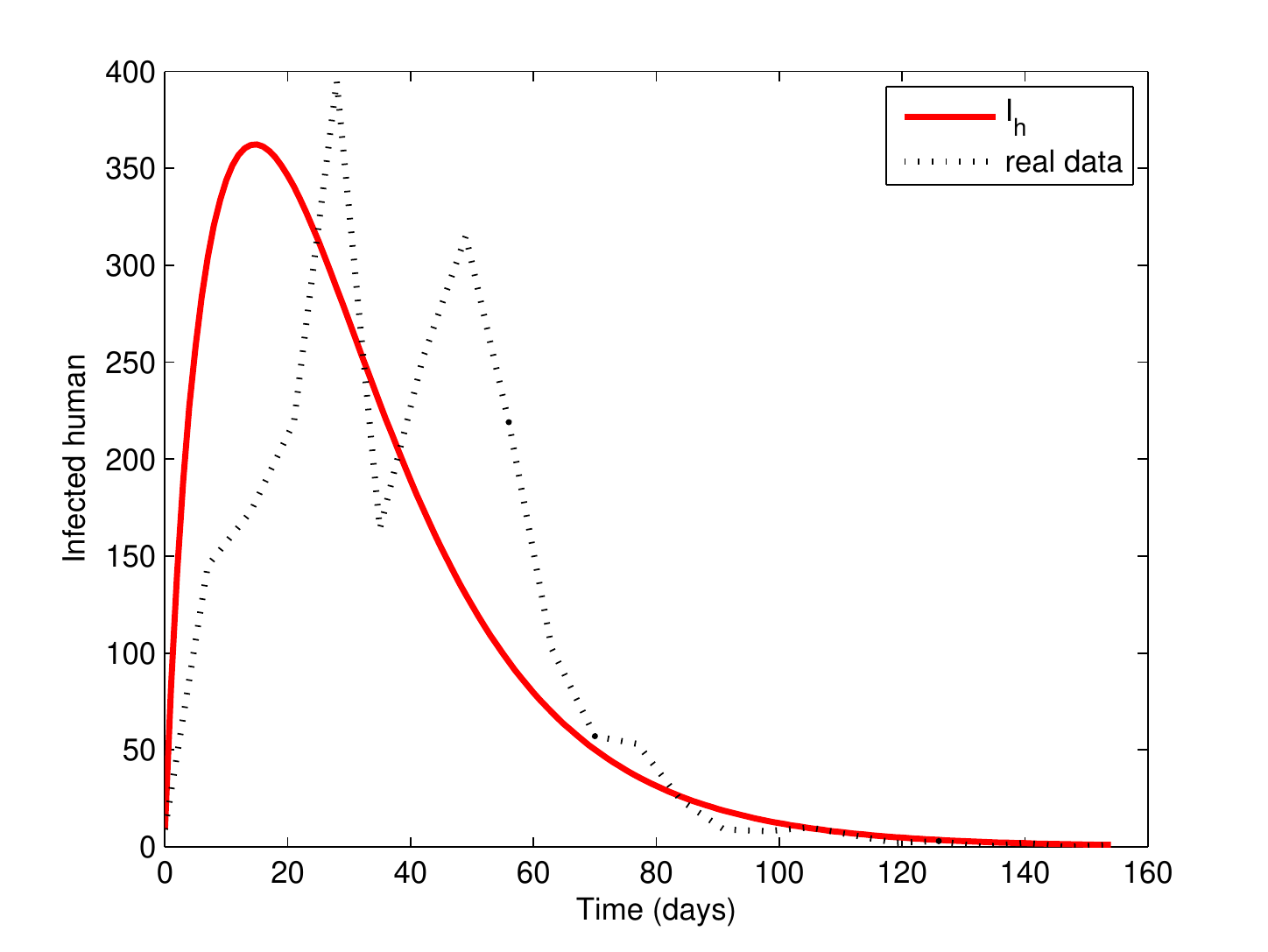}
\caption{(F) $c_m \equiv 0.02$}
\label{madeira_insecticida_2}
\end{subfigure}
\begin{subfigure}[b]{0.48\textwidth}
\centering
\includegraphics[width=\textwidth]{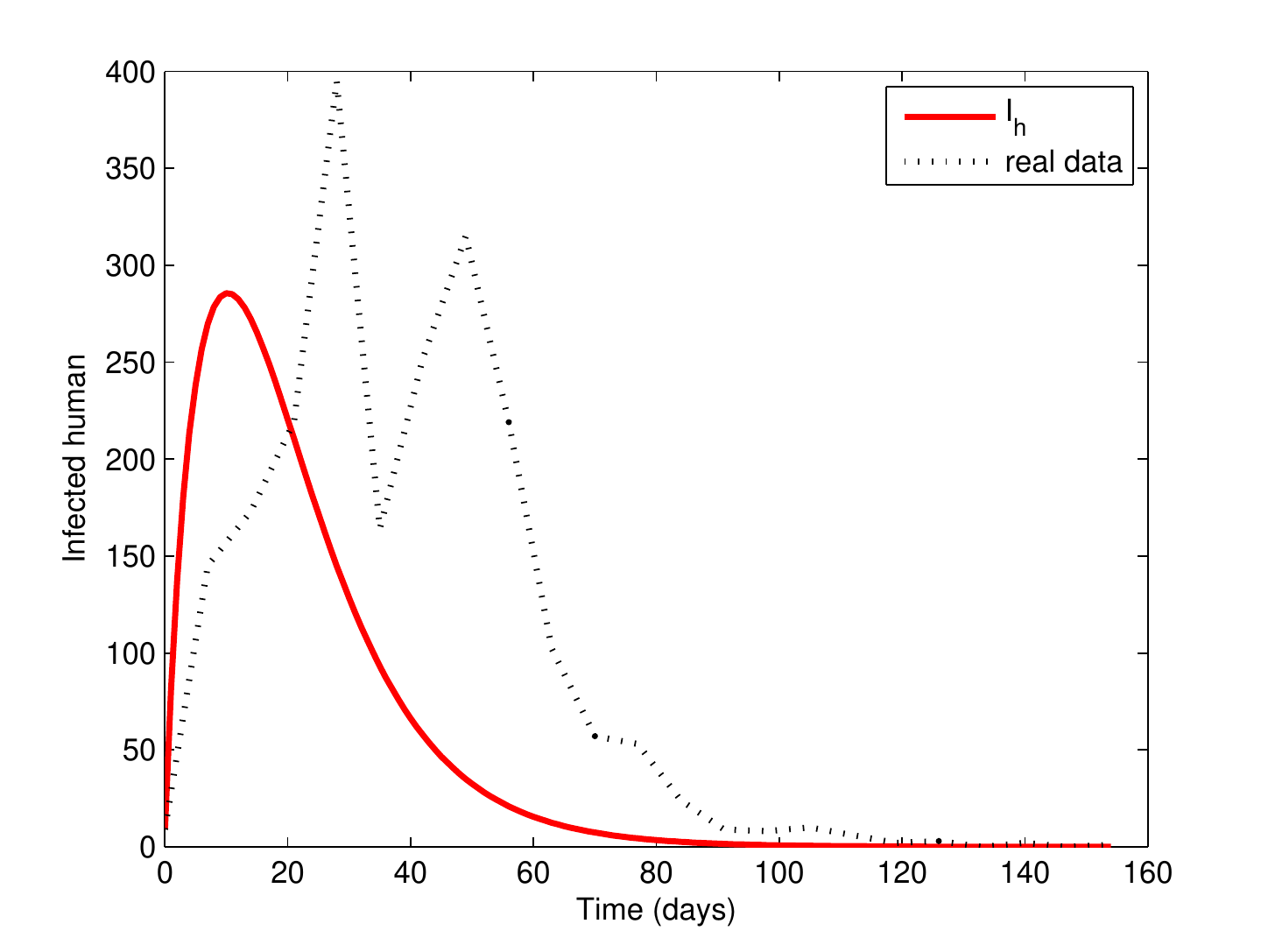}
\caption{(G) $c_m \equiv 0.05$}
\label{madeira_insecticida_5}
\end{subfigure}
\caption{Number of infected individuals for a constant
educational campaign of $1-\alpha \equiv 0.5$ and distinct levels
of insecticide (continuous line) \emph{versus} observed real data (dotted line)}
\label{insecticide}
\end{figure}

In this section, numerical results are presented. Our aim is to show a simulation
of the possible evolution of the dengue outbreak occurred on Madeira island, using
parameterized and validated epidemiological and entomological data.
The human data was adapted to the Madeira region through official data \cite{DGS,censos2011}.
Despite the research efforts, some information required for the model
parametrization still lacks, especially entomological. This is due to the difficulty of obtaining
it by laboratory assays. Even when experiments are possible, sometimes the mosquito
behavior presents distinct features when in a controlled environment or in nature
\cite{Luz2003}. For this reason, a range of values for mosquito parameters where analyzed
(see Table~\ref{table_1} for details). The initial values for the system
of differential equations \eqref{cap6_ode1}--\eqref{cap6_ode2} are:
\begin{gather*}
S_h(0)=111991, \quad  I_h(0)=9, \quad R_h(0)=0,\\
A_m(0)=111900 \times 6, \quad S_{m}(0)=111900 \times 3, \quad I_m(0)=1000.
\end{gather*}
The software used in the simulations was \texttt{Matlab}, with the routine \texttt{ode45}.
This function implements a Runge--Kutta method with a variable time step
for efficient computation. Figure~\ref{educational} shows different simulations
for educational campaigns, without application of insecticide.
This was the major control measure to fight the disease.
It is possible to see that educational campaigns have an important role
in the decrease of infected human and the best curve that fits the real data has an
implementation of ecological control between $50\%$ and $75\%$.
Table~\ref{table_3} presents the total number of infected human individuals for the simulations done.
Without any control measure, about 12\% of all population of Funchal would be infected.
A common sense conclusion is that when we increase the proportion of control measures, the
number of infected decreases considerably. In the same manner, the application of
even small quantities of insecticide, seems to increase the effects of ecological control
(compare simulations $C$ and $E$). The explanation for this lies in the fact that, when
an outbreak occurs, the application of an efficient adulticide will immediately affect
the transmission rate of the virus. Educational campaigns,
even being a good strategy for the ecological control
of the mosquito, imply time to promote the
necessary motivation for the people to react to the disease.
\begin{table}[ptbh]
\begin{center}
\caption{Total number of infected individuals}
\label{table_3}
\begin{tabular}{c c c}
\hline
Simulations & Control Values & Total Number of Infected\\
\hline
A & no control, i.e. $1-\alpha = c_m \equiv 0$ & 13677\\
B & $1-\alpha \equiv 0.25$ and $c_m \equiv 0$  & 7719\\
C & $1-\alpha \equiv 0.50$ and $c_m \equiv 0$  & 4827\\
D & $1-\alpha \equiv 0.75$ and $c_m \equiv 0$  & 3388\\
\hline
E & $1-\alpha \equiv 0.5$ and $c_m \equiv 0.01$ & 3073\\
F & $1-\alpha \equiv 0.5$ and $c_m \equiv 0.02$ & 2210\\
G & $1-\alpha \equiv 0.5$ and $c_m \equiv 0.05$ & 1179\\
\hline
real data &  & 2168\\
\hline
\end{tabular}
\end{center}
\end{table}
The graphs with education campaigns at 50\% and a variation of insecticide
application are shown in Figure~\ref{insecticide}.
The curves that better illustrate the peak of the epidemics use between
0\% to 1\% of insecticide. However, when compared to
the table of total infected cases (Table~\ref{table_3}),
the nearest simulation is F ($\alpha=50\%$ and $c_m=2\%$).
In fact, this difference can be explained by the fact that the simulation is
made by comparing the total infected cases with the notified ones. It is well known by
Health Authorities that, besides the asymptomatic cases,
some patients do not go to the Health Centers: not only
because they have light symptoms but also because
their relatives or neighbors had already had dengue
and they think they can handle the situation by themselves, at home.

\begin{remark}
A simple tuning of the control parameters $\alpha$ and $c_m$,
by using some optimization technique like least square or curve fitting,
does not seem appropriate here. Using the least square method
to choose the best combination of the two controls,
we obtained the proportion of adulticide
$c_m=0.0280$ and the value for educational campaigns $1-\alpha=0$.
This last value implies that the ecological control
has no influence whatsoever in the system, which
is not in agreement with the case under study.
\end{remark}


\section{Conclusions}
\label{Sec:6}

One of the most important issues in epidemiology
is to improve control strategies with the final goal
to reduce or even eradicate a disease.
In this paper a dengue model based on two populations,
humans and mosquitoes, with educational and
insecticide control measures, has been presented.
Our study provides some important epidemiological insights about the impact
of vector control measures into dengue in Madeira island.
The work was done with collaboration of the \emph{Instituto de Higiene e Medicina Tropical},
which provided us with valuable information about the disease characteristics and entomologic aspects,
and \emph{Instituto de Administra\c{c}\~{a}o da Sa\'{u}de e Assuntos Sociais} from Madeira,
which gave us specific information about the outbreak, namely real numbers of the disease,
affected areas and what kind of control was done in the island. Such cooperation
and discussions with entomologists and doctors, was crucial to tune the parameter values
of the mathematical model. Our results show how dengue burden can decrease
with the help of vector control measures such as insecticide and ecological control.
We concluded that small quantities of insecticide have a considerable impact in the
short time intervention when an outbreak occurs. The application of educational campaigns
decreases the disease burden and can act as a long time prevention.
As future work, we intend to add an optimal control analysis
to decide whether a given combination of control values is the best.
Such analysis will be important for policy makers to know the
optimal combination of the control strategies.


\section*{Acknowledgements}

This work was partially supported by The Portuguese Foundation
for Science and Technology (FCT): Rodrigues and Torres through
the Center for Research and Development in Mathematics and Applications (CIDMA)
within project PEst-OE/MAT/UI4106/2014; Monteiro by the ALGORITMI Research Centre
and project PEST-OE/EEI/UI0319/2014; Silva and Sousa by the project
``Dengue in Madeira archipelago. Risk assessment for the emergence
of Aedes aegypti mediated arboviroses and tools for vector control''
with reference PTDC/SAU-EPI/115853/2009.
The authors are very grateful to two anonymous referees,
for valuable remarks and comments, which
contributed to the quality of the paper.



\end{document}